\documentclass{article}
\usepackage{dcase2024,amsmath,graphicx,url,times,booktabs,tabularx}
\usepackage{enumitem,xcolor}
\usepackage{hyperref}
\hypersetup{colorlinks=true}
\usepackage{booktabs}
\usepackage{multirow}
\usepackage{tabularx}

\title{Heterogeneous sound classification with  \\the \emph{Broad Sound} Taxonomy and Dataset}

\name{
  Panagiota Anastasopoulou,
  Jessica Torrey,
  Xavier Serra,
  Frederic Font
}

\address{
  Music Technology Group, Universitat Pompeu Fabra, Barcelona, Spain \\
  {\fontsize{11}{10}\selectfont
  panagiota.anastasopoulou@upf.edu, 
  jessica@jessicatorrey.com, 
  xavier.serra@upf.edu,
  frederic.font@upf.edu
  }
}

\begin{document}

\ninept
\maketitle

\begin{sloppy}

\begin{abstract}
Automatic sound classification has a wide range of applications in machine listening, enabling context-aware sound processing and understanding. This paper explores methodologies for automatically classifying heterogeneous sounds characterized by high intra-class variability. 
Our study evaluates the classification task using the Broad Sound Taxonomy, a two-level taxonomy comprising 28 classes designed to cover a heterogeneous range of sounds with semantic distinctions tailored for practical user applications.
We construct a dataset through manual annotation to ensure accuracy, diverse representation within each class and relevance in real-world scenarios. 
We compare a variety of both traditional and modern machine learning approaches to establish a baseline for the task of heterogeneous sound classification. 
We investigate the role of input features, specifically examining how acoustically derived sound representations compare to embeddings extracted with pre-trained deep neural networks that capture both acoustic and semantic information about sounds. 
Experimental results illustrate that audio embeddings encoding acoustic and semantic information achieve higher accuracy in the classification task. 
After careful analysis of classification errors, we identify some underlying reasons for failure and propose actions to mitigate them. 
The paper highlights the need for deeper exploration of all stages of classification, understanding the data and adopting methodologies capable of effectively handling data complexity and generalizing in real-world sound environments.
\end{abstract}


\section{Introduction}
\label{sec:intro}

Sound classification plays a crucial role in numerous applications ranging from sound and music analysis, browsing and retrieval to acoustic monitoring and ubiquitous computing~\cite{font2018sound}. 
Automatic analysis of diverse sound types necessitates the extraction of relevant features from audio signals, combined with machine learning techniques.
This has garnered significant attention from fields focused on music, speech, and environmental sounds, leading to the development of various taxonomies and algorithmic techniques tailored to different applications.

In this paper, we concentrate on a general-purpose classification framework where, instead of focusing on a particular type of sound, the goal is to classify \emph{any} type of input sound.
For that purpose, we previously developed the Broad Sound Taxonomy (BST), which organizes sounds into a two-level hierarchical structure with 5 top-level and 23 second-level classes~\cite{anastasopoulou2024general}.
The top level of the taxonomy consists of the classes \emph{Music}, \emph{Instrument samples}, \emph{Speech}, \emph{Sound effects}, and \emph{Soundscapes}. 
A diagram with all classes (and their abbreviated names) can be seen in Fig.~\ref{fig:taxonomy}.
The taxonomy is designed to be user-friendly and accommodates a wide diversity of sounds, ensuring the classes are easy to understand, broad, and comprehensive.
These classes exhibit significant intra-class variability, primarily influenced by the semantic foundation upon which the taxonomy was constructed.
Such intra-class variability means that sounds of the same class can exhibit very different acoustic characteristics.
Our goal is to build a sound classification system that can successfully classify sounds using the BST taxonomy. 
To that end, we curate a dataset comprising 10k sounds annotated with the BST classes. 
We use k-NN classifiers and study their performance using input sound representations that capture different levels of acoustic and/or semantic information. 
Besides the classifier performance metrics, we conduct manual error analysis and systematically characterize the model's misclassifications.
The moderate number of classes in the taxonomy proves advantageous in this step, enabling easier human evaluation of algorithmic mistakes. 
By analyzing misclassifications, we are able to suggest ways in which the classification system can be further improved.

The proposed approach and findings have broad applicability, as the automatic extraction of the systematized knowledge from such a hierarchical structure can streamline the organization, annotation, and retrieval of audio data, along with other related tasks across diverse domains.
Using such a classifier, capable of categorizing any type of sound into broad categories, can be useful for providing an initial context of a sound class and thereafter for carrying out context-aware processing of sounds.

\begin{figure}[t]
  \centering
\centerline{\includegraphics[width=\columnwidth]{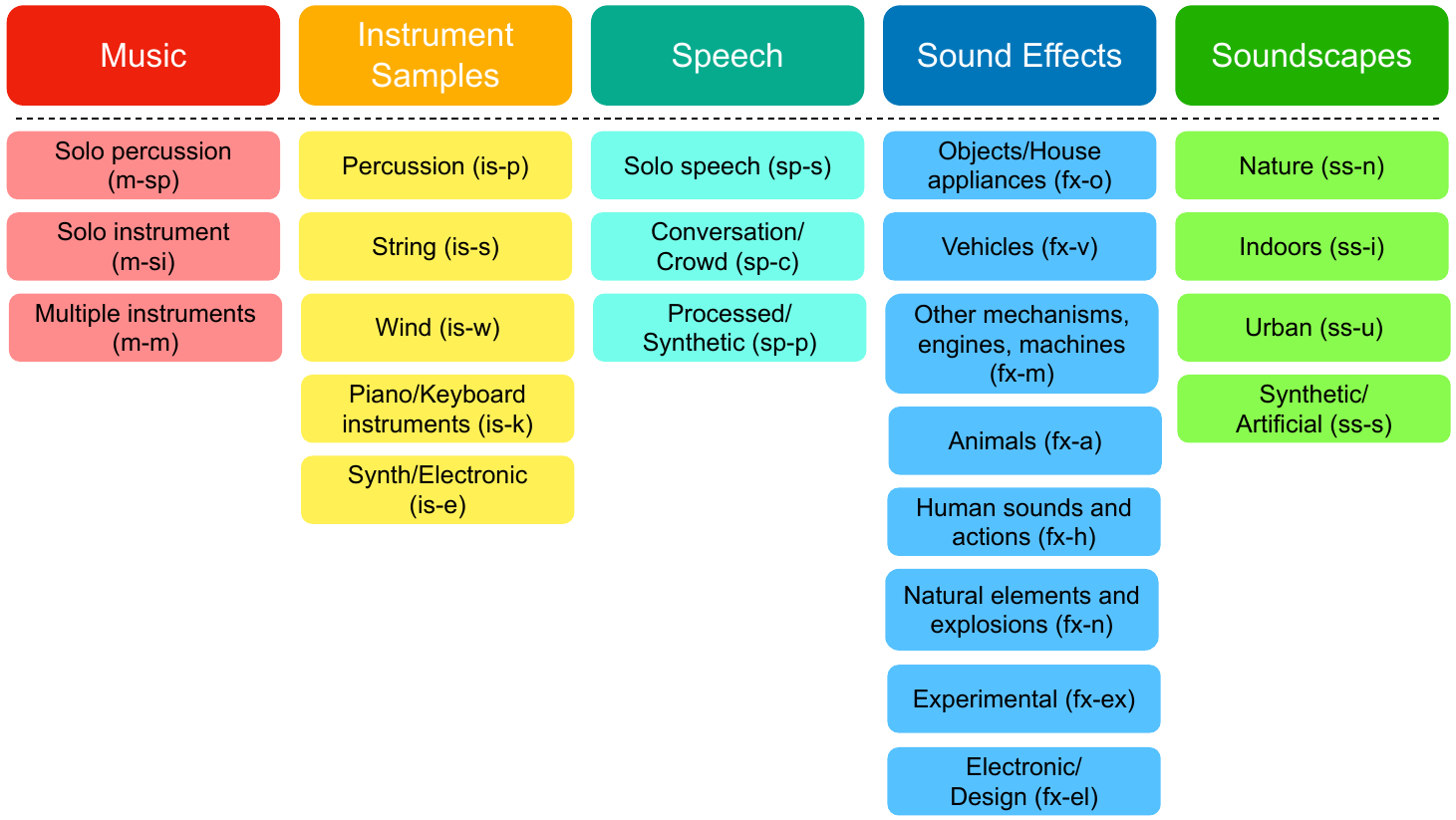}}
  \caption{Class hierarchy for the Broad Sound Taxonomy (BST).}
  \label{fig:taxonomy}
\end{figure}

\section{Background}
\label{sec:back}

Over the years, several taxonomies have been proposed for organizing sound. 
Most taxonomies are tailored to specific domains or tasks, as exemplified by works on sound design~\cite{moffat2017unsupervised, nielsen2020ucs}, urban or environmental scene analysis~\cite{piczak2015esc,salamon2014dataset,lafay2019investigating} and music or instrument categorization~\cite{defferrard2016fma,ramires2020loop,engel2017nsynth}, while other taxonomies are designed to cover general use cases (e.g. Google's AudioSet~\cite{gemmeke2017audioset}).
On the one hand, when existing taxonomies are \emph{simple} (i.e.~low number of classes with shallow hierarchy), they tend to be domain-specific and are not comprehensive enough to generally classify \emph{heterogeneous} sounds (e.g. ESC-50~\cite{piczak2015esc}, Urban Sound Taxonomy~\cite{salamon2014dataset}, FMA~\cite{defferrard2016fma}, NSynth~\cite{engel2017nsynth}).
On the other hand, general-purpose taxonomies are often very complex or lack a user-centric design (e.g. AudioSet has over 500 sound classes organized in a deep hierarchy), meaning that only expert users can use them effectively.
The aforementioned Broad Sound Taxonomy addresses the lack of a simple yet comprehensive sound taxonomy that can be easily understood and used by sound practitioners of different levels of expertise and, at the same time, provide informative sound classes relevant to various applications such as sound analysis and retrieval~\cite{anastasopoulou2024general}.

In the field of machine listening, automatic sound classification has been typically addressed using machine-learning classifiers such as k-Nearest Neighbors (k-NNs), Support Vector Machines (SVMs), Multilayer Perceptrons (MLPs), and Hidden Markov Models (HMMs)~\cite{virtanen2018computational}.
These classifiers traditionally rely on features such as Mel-frequency cepstral coefficients (MFCCs) and other spectrum-based representations that only capture acoustic information of sounds.
In recent years, different types of deep neural networks (DNNs) have gained prominence across the audio field due to their superior performance. 
One notable use is their ability to effectively transform raw audio data into highly meaningful representations.
Because such representations are often obtained from models trained on classification tasks, they do not only capture acoustic information about sounds, but also encode some level of semantic information.
Models such as VGGish\cite{hershey2017cnn}, YAMNet~\cite{yamnet}, or FSD-SINet~\cite{fonseca2021improving}, produce high-level, semantically meaningful embeddings while using audio as input.
Another recent approach is the use of contrastive learning techniques to train models that learn a joint audio and language embedding space in which sound semantics are even more prominent.
An eminent example is the CLAP architecture~\cite{elizalde2023clap,wu2023large}, which learns audio concepts from natural language sound descriptions.
These learned feature representations can be used as input features with traditional machine learning classifiers for addressing downstream tasks, which is typically known as \emph{transfer learning}. 
Through transfer learning, less complex models can efficiently leverage pre-trained models to achieve high accuracies in downstream tasks~\cite{alonso2022music,sanguineti2020leveraging,eck2007automatic}.
In this work, we use transfer learning to address the task of heterogeneous sound classification.

\section{Methodology}
\label{sec:method}

\subsection{Dataset creation}
\label{subsec:datset}

We introduce the Broad Sound Dataset (BSD), a collection of annotated sounds aligned with the second level of the classes defined in the BST taxonomy (Fig.~\ref{fig:taxonomy}).
The initial release, a contribution of this paper, contains more than 10,000 sounds and is named BSD10k.
BSD10k has been built using sounds obtained from Freesound, a website that hosts over 650,000 diverse sounds released under Creative Commons (CC) licenses and contributed by a wide range of individuals~\cite{font2013freesound}.
We leveraged existing public Freesound-based datasets such as FSD50K~\cite{fonseca2021fsd50k}, freefield1010~\cite{stowell2013open}, Freesound Loop Dataset~\cite{ramires2020loop}, together with other in-house Freesound collections to compile an initial list of approximately 60,000 sound candidates of heterogeneous nature. 
These candidates were assigned to one of the five top-level classes of the BST taxonomy by leveraging their ground-truth labels from their original dataset(s) and using other heuristics based on basic signal processing techniques (e.g. onset detection) and available Freesound metadata (e.g. sound tags).
After mapping the candidates to the top level of the taxonomy, a manual annotation phase was carried out to address potential inaccuracies and assign the corresponding second-level taxonomy category to each sound candidate.

For the annotation phase, we developed an in-house online annotation tool which was used by the authors of the paper to get familiar with the taxonomy and carry out the annotations.
For each candidate sound, the annotators selected the most appropriate second-level class and provided a confidence level for their annotation. 
The provided confidence level is not used for the classification tasks in this paper, but it helps ensure a more accurate annotation process and may provide useful data for future experiments~\cite{mendez2022eliciting}.
The original sound title and tags from Freesound were presented to the annotators to facilitate the annotation of acoustically ambiguous sounds.
During the course of three months, the annotators classified 10,309 sounds, resulting in a total duration of 32.5 hours of audio, which forms the final BSD10k dataset.
The annotated data has a non-uniform class distribution, leading to data imbalance, with some classes having over 1,000 sounds while others are represented by approximately 100 sounds. 
The top-level division of the audio data is 1635 \emph{Music}, 2094 \emph{Instrument samples}, 1250 \emph{Speech}, 3911 \emph{Sound effects}, and 1419 \emph{Soundscapes}.

The Freesound audio data is heterogeneous, not only in content but also in quality, devices of recording, and lengths. 
Even though many sounds use (semi-)professional recording equipment~\cite{fonseca2021fsd50k}, this diversity can be used as an advantage in developing a general-purpose classifier that generalizes well.
During the annotation, we also monitored the diversity within each class; e.g. in the \emph{Natural sounds and explosions} class, we ensured the presence of water sounds, rocks, as well as lightning and fireworks.
The length of the sounds also varies, following a U-shape distribution.
Longer samples were cropped to a maximum of 30 seconds, as sounds of this nature —often music or soundscapes— tend to repeat information.
Even though we start with candidates from existing datasets, we download the original files using their IDs from the Freesound API.
We then transform all sounds to adhere to a standardized format of uncompressed 44.1 kHz 16-bit mono audio files.
The dataset is released with an open license and is publicly accessible\footnote{\url{https://github.com/allholy/BSD10k}}.

\subsection{Sound representations}
\label{ssec:feature}

We compare a selection of different types of sound representations, which are chosen to capture distinct levels of acoustic and semantic features. 

\begin{description}[leftmargin=0.7cm]
\item[FSSimRep:]
We extract a feature representation derived from various spectral, time-domain, rhythm and tonal characteristics calculated using signal-processing algorithms with the FreesoundExtractor\footnote{\url{https://essentia.upf.edu/freesound_extractor.html}} of the Essentia audio analysis library~\cite{bogdanov2013essentia}.
With an audio file given as input, the FreesoundExtractor provides several statistics for each of the features above, which are then aggregated into a vector of $846$ dimensions and scaled to be in the range $[0, 1]$. 
The scaled vector is reduced to $100$
dimensions using Principal Component Analysis (PCA), producing the final sound representation.
This representation is akin to the representation currently used for the sound similarity feature in Freesound, and it is expected to only capture the acoustic properties of sounds.

\item[VGGish and FSD-SINet:]
We utilize the embeddings from VGGish~\cite{hershey2017cnn} and FSD-SINet~\cite{fonseca2021improving}. 
They are both large convolutional neural network (CNN) models trained on audio signals in classification tasks.
These models take audio signals as input and are expected to learn both about their acoustic properties and semantic meaning by relating audio signals to the classification labels.
We use both models as two examples of classification-based embeddings trained on distinct datasets (YouTube100M and FSD50K), with output representation dimensions of \((n, 128)\) and \((n, 512)\), respectively, where $n$ represents the number of frames dependent on the length of the audio file. 
To obtain the final one-dimensional vector representation, we carry out temporal aggregation by averaging over $n$ frames.

\item[LAION-CLAP:]
Finally, in our experiments, we include embeddings extracted from the multi-modal LAION-CLAP model~\cite{wu2023large}.
CLAP uses contrastive learning techniques to acquire knowledge from pairs of audio signals and natural language textual descriptions. 
This approach allows the model to be fed not only with the audio signals but also with rich contextual semantic information about them. 
Given an audio file as input, LAION-CLAP provides a final $512$-dimensional vector representation.
\end{description}

\subsection{Model and evaluation metrics}
\label{ssec:model}

For our experimental setup, we use the k-Nearest Neighbors (k-NN) algorithm 
as our classifier. 
The choice is motivated by its low complexity, interpretability, and common use in transfer learning settings. 
To complement our experiments, we run preliminary experiments using Support Vector Machine (SVM) models and obtained results similar to those reported by k-NN models, therefore we will not report SVM results in this paper.

To identify the optimal hyperparameters, we compare various sets of model parameters to determine the most effective configuration for model performance. 
We conduct a grid search to systematically explore the hyperparameter space, evaluating different numbers of neighbors, distance metrics, and weighting schemes~\cite{yang2020hyperparameter}.
To evaluate the performance of the trained models, we calculate accuracy, precision, recall and F1-score evaluation metrics.
We divide our dataset into two splits used for training and evaluation. The evaluation split consists of a random selection of 40 sounds for each second-level class of the taxonomy, totaling 920 sounds (\~9\% of the size of the dataset). 
We assess qualitatively that the random selection for the evaluation set resulted in high intra-class sound variations.
The rest of the sounds are included in the training set. 

Additionally, we take advantage of the hierarchical structure of the taxonomy to run experiments using only the top-level classes as labels, grouping sounds with similar semantics and reducing the total number of classes to five (Fig.~\ref{fig:taxonomy}).
For consistency, we use the same data split for the top-level training process.
Although this approach introduces imbalance in the number of test samples per class due to the varying number of second-level classes within each top-level class, it ensures a fair comparison in the evaluation process.

To obtain further insights about classification performance, we characterize the errors by manually reviewing all misclassified sounds from the best-performing model across all input representations, as well as 200 randomly sampled misclassifications from the best models of the remaining input representations. 
This analysis is performed for both second-level and top-level classification setups. 
We identify the potential reasons for each misclassification and then consolidate the most common reasons into error categories.


\section{Results}
\label{sec:results}

\begin{table}[t]
  \centering
  \caption{Accuracy and F1-score for the best-performing k-NN per input sound representation.}
  \label{tab:metrics}
  {
    \setlength{\tabcolsep}{4pt}
    \begin{tabular}{lcccc}
      \toprule
      \multirow{2}{*}{\textbf{Model input}} & \multicolumn{2}{c}{\textbf{Second-level}} & \multicolumn{2}{c}{\textbf{Top-level}} \\
      & \textbf{Accuracy} & \textbf{F1-score} & \textbf{Accuracy} & \textbf{F1-score} \\
      \midrule
      FSSimRep & 0.426 & 0.40 & 0.678 & 0.667 \\
      VGGish & 0.527 & 0.506 & 0.748 & 0.741 \\
      FSD-SINet & 0.562 & 0.544 & 0.746 & 0.746 \\
      LAION-CLAP & \bf{0.761} & \bf{0.748} & \bf{0.873} & \bf{0.868} \\
      \bottomrule
    \end{tabular}
  }
\end{table}

\subsection{Performance metrics}
Table~\ref{tab:metrics} shows the classification accuracies and F1-scores of the k-NN classifiers trained with the different input representations we compare. 
We report the accuracy and F1-score of the best-performing classifier for each input representation according to the hyperparameter optimization.
We observe that, in almost all instances, the highest recall coincides with the highest accuracy.
This suggests that comparing accuracies across various input representations, including the top-level classifiers with an unbalanced test set, remains a reliable metric without inherent bias towards classes with larger sample sizes.

Both accuracy and F1-score metrics show that classification performance improves when classifying at the top level compared to the second level of the taxonomy (average of 0.19 for accuracy and 0.21 for F1-score). 
This is expected as the task becomes progressively easier with fewer number of classes, introducing greater orthogonality at the first level of the taxonomy.
The increase in performance comparing top-level with second-level is significantly lower for CLAP (0.11 for accuracy and 0.12 for F1-score), which could be attributed to the fact that CLAP captures sound semantics more efficiently and therefore it can perform better in the second-level of the taxonomy where class semantics are more nuanced.

The CLAP embeddings outperformed the other representations in both top-level and second-level classification tasks.
This suggests that the joint audio-language embedding space captures acoustic and semantic information better, which is beneficial for the classification of heterogeneous sounds.
VGGish and FSD-SINet result in very similar performances.
Despite our expectation that FSD-SINet would outperform VGGish due to its training on the FSD50K dataset, which includes Freesound data relevant to our task, 
both models show comparable results.
They have an average of 0.216 and 0.223 lower than CLAP for accuracy and F1-score, respectively. 
This suggests that these embeddings do not capture acoustic and semantic sound properties with the same richness as CLAP embeddings. 
Finally, the models trained with FSSimRep exhibit the lowest performance, an average of 0.101 and 0.106 lower than VGGish for accuracy and F1-score, respectively. 
This highlights the challenge of distinguishing classes using solely acoustic information due to intra-variability and acoustic diversity within classes.

Fig.~\ref{fig:cm} shows the confusion matrix for the best-performing k-NN model trained with the CLAP sound representation and holds insights into how the model performs for each individual second-level class of the taxonomy.
We observe that most classes exhibit very good performance, yet there are instances of lower performance in specific classes, such as \emph{Conversation/Crowd}.
This discrepancy may stem from factors such as data imbalance, class complexity, or reduced orthogonality between certain classes. 

\begin{figure}[t]
  \centering
\centerline{\includegraphics[width=1.0\columnwidth]{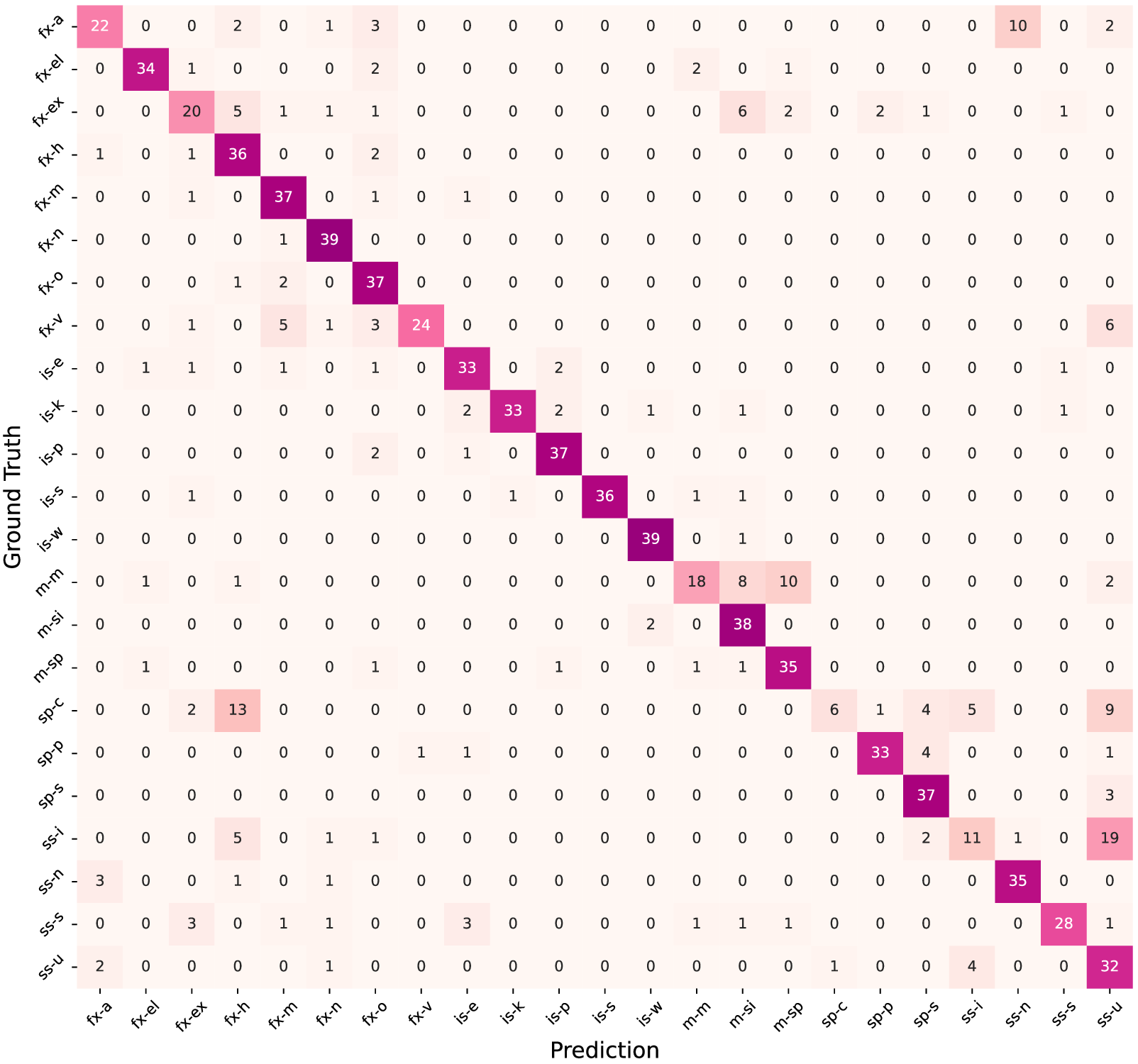}}
  \caption{Confusion matrix for the best-performing k-NN model trained with CLAP.}
  \label{fig:cm}
\end{figure}

Regarding hyperparameter optimization, we find that the variation in accuracy among the top 100 grid search configurations for each embedding training remains small, with a maximum difference of approximately 0.065 (and 0.035 for top classes). 
The top 100 choices include nearly all neighbors, distance metrics, and weighting schemes, indicating stable performance across a broad area of the hyperparameter space.
This stability suggests that specific hyperparameters have little impact on the performance of this task when leveraging embeddings, regardless of their efficacy.

\subsection{Error characterization}
\label{ssec:error}

Table~\ref{tab:errors} shows the results of the error characterization. 
We observe that the most common reason for the misclassification of sounds is when sounds fall ambiguously \emph{between classes}, either between second-level classes with a common top-level class (14.6\%), or between second-level classes belonging to a different top-level class (26\%). 
That suggests that even humans may have difficulty distinguishing these classes. 
Further insights about that matter could be obtained by analyzing the confidence annotation scores included in BSD10k.
We also observe that simplifying the task in the top-level classification does not significantly reduce \emph{between classes} errors.
Interestingly, errors are more prevalent between different top-level categories than within the same one, indicating potential for enhancing the classifier's capability to differentiate between higher-level classes to improve overall hierarchical classification accuracy.
Analyzing the discrepancies between the top-level and second-level classifiers reveals that $54\%$ of errors across all second levels are accurately predicted by the top-level classifier, supporting the claim that integrating hierarchical information within a unified model is a promising future direction.
Additionally, a notable portion of these errors are linked to the lowest-performing class (\emph{Conversation/Crowd}), suggesting that improving the dataset or model to better handle less orthogonal classes could lead to better overall results.

Misclassifications due to \emph{common source} (i.e classes include sounds from the same source), \emph{single-source evolution} (i.e sound from one source evolves over time), or \emph{prominence of one source} (i.e. one sound dominates in duration or loudness) are influenced by the taxonomy's nature, which separates sound samples even when they originate from the same source (e.g. birds as part of a soundscape \emph{vs} isolated birds, or human talking \emph{vs} human crying).
Because of the class definitions, the model is tasked to learn deeper semantic distinctions and information about the source mixture, thereby making the classification task more complex.
To reduce these errors, models could integrate mixture and context-aware learning strategies during training. 
Errors grouped under \emph{acoustic ambiguity} have one or more acoustic properties that resemble another sound from a different class (sounds \emph{like} x, \emph{is} y).
Emphasizing semantic information could mitigate these errors, as they are more pronounced in the lower-performing models with less semantic integration, constituting $43-54\%$ of their total errors (against $23-27\%$ for CLAP).
We note, though, that confusing sounds with very high acoustic similarity may be less consequential in certain tasks, such as sound design.

\begin{table}[t]
    \centering
    \caption{Error characterization for the best-performing k-NN model trained with CLAP.}
    \label{tab:errors}
    \vspace{2pt}
    {
    \begin{tabular}{lcc}
        \toprule
        \textbf{Error category} & \textbf{\footnotesize Second-level} & \textbf{\footnotesize Top-level} \\ 
        \midrule
        Acoustic ambiguity & 60 & 27\\ 
        Between classes (different top) & 57 & 52\\ 
        Between classes (same top) & 32 & - \\
        Common source & 18 & 10 \\ 
        Prominence of one source & 23 & 18\\ 
        Single-source evolution & 3 & 2 \\ 
        Low quality & 3 & 0 \\ 
        Uncommon/Weird/Other & 24 & 8\\ 
        \textbf{Total} & \textbf{220} & \textbf{117} \\ 
        \bottomrule
    \end{tabular}
    }
\end{table}

\section{Conclusions}
\label{sec:conclusion}

In this paper, we present a comparative analysis of various input representations with different levels of acoustic and semantic information for the task of heterogeneous sound classification. 
To address the challenges posed by the classification of a broad taxonomy with significant intra-variability, we introduce the manually curated BSD10k dataset which enables automatic classification tasks and offers valuable data pools for diverse research tasks.
To baseline the problem and understand the error margin, we complement the evaluation metrics with manual error characterization through auditory evaluation of the misclassifications.
Our findings indicate that greater semantic information enhances classification performance and insertion of hierarchical information during training can prove beneficial.
Organizing available data into simpler taxonomic structures can improve the sound description process and enable the training of reliable automatic classifiers, providing a pre-processing step for context-aware sound processing and understanding.

\section{ACKNOWLEDGMENT}
\label{sec:ack}
This research was partially funded by the Secretaria d'Universitats i Recerca del Departament de Recerca i Universitats de la Generalitat de Catalunya (ref. 2023FI-100252) under the Joan Oró program, and the IA y Música: Cátedra en Inteligencia Artificial y Música (TSI-100929-2023-1) by the Secretaría de Estado de Digitalización e Inteligencia Artificial and the European Union-Next Generation EU under the program Cátedras ENIA 2022.

\bibliographystyle{IEEEtran}
\bibliography{refs}

\end{sloppy}
\end{document}